# IoT Security and Authentication schemes Based on Machine Learning: Review

[1]Rushit Dave
*Assistant Professor of Computer Science*
*University of Wisconsin Eau Claire, USA*

---

---

**ABSTRACT:** With the latest developments in technology, extra and extra human beings depend on their private gadgets to keep their touchy information. Concurrently, the surroundings in which these gadgets are linked have grown to grow to be greater dynamic and complex. This opens the dialogue of if the modern-day authentication strategies being used in these gadgets are dependable ample to preserve these user's records safe. This paper examines the distinct consumer authentication schemes proposed to make bigger the protection of exceptional devices. This article is break up into two one of a kind avenues discussing authentication schemes that use both be- havioral biometrics or physical layer authentication. This survey will talk about each the blessings and challenges that occur with the accuracy, usability, and standard protection of computing device getting to know strategies in these authentication systems. This article targets to enhance in addition lookup in this subject via exhibiting the more than a few present day authentication models, their schematics, and their results.

## I. INTRODUCTION

Within the closing few decades, developments in science are taking place at an increasingly quickly fee and units are swiftly turning into greater various and complex. Concurrently, the quantity of touchy data being saved on non-public units has additionally considered a dramatic rise, as nicely as the chance of this records being compromised. There are a multitude of approaches to authenticate a single user, with the most frequent approach being static methods, i.e., passwords and Personal Identification Numbers (PINs). Although reliable, as soon as a password or PIN is compromised, that account is prone to every person who has received get entry to that touchy data. To counter this problem, researchers have been investigating the legitimacy of dynamic authentication systems. With the aforementioned shortcomings of static authentication, dynamic authentication provides a broad range of benefits, such as non-stop authentication, improved flexibility, and in some cases, much less labor for the user. This survey will be exploring two often pro- posed structures for dynamic authentication and their efficacy; these two areas being physical-layer authentication (PLA) and biometric-based authentication methods.

Biometrics are a user's bodily or behavioral characteristics, these of which can be used in person authentication. Biometrics are classified into two categories: physiological and behavioral. Physiological biometrics consists of scanning bodily elements of the user, for instance the usage of a user's palm prints for authentication [1]. On the different hand, behavioral biometrics focuses on now not a password, or a bodily trait of the user, however the user's behaviors in how they engage with the object wanting authentication. Many of these biometrics have already been applied into telephones and clever units for the previous couple of years, however they are additionally being used in different time-honored structures such as the healthcare device to amplify their safety [2].

However, physiological biometrics have determined to be no longer as in your price range and no longer as correct as researchers hoped for. Physiological biometrics require extra scanners and software program to be hooked up onto the gadget so that the machine can scan the user's bodily facets [3]. This sooner or later ends up growing the fee to produce the product and additionally increasing the value to purchase the product. This poses the question: is there a higher gadget to restoration the issues related with physiological biometrics? This underperforming authentication scheme has brought about researchers to seem for a greater environment friendly option, and that effectivity is discovered the usage of behavioral biometrics in authentication schemes. Behavioral biometrics solves the trouble of fee due to the fact the biometrics do now not require any extra software program to be hooked up into the gadgets in order to scan behavioral features. Research has been con- ducted the usage of these behavioral biometrics and exclusive computing device studying algorithms in hopes to make our gadgets extra impervious and extra efficient.

Based on preceding predictions made in 2018, an estimated 50 billion units are presently linked to the web [4,5]. Most of the gadgets related to the net are now not standard computers, laptops, or smartphones, however, alternatively are gadgets that are smartwatches, clever refrigerators, thermostats, voice services, protection cameras, and lots more. These gadgets have been given more than a few applied sciences to talk with the web to supply higher performance and effectivity for the end-user and they make up what is regarded the Internet of Things. One of the defining traits of the Internet of Things is the interconnectivity between its devices. Gateway gadgets are accountable for accumulating information from surrounding sensor gadgets and transferring it to the internet. Both the gateway and sensor gadgets need to be secured, and measures want to be in vicinity to forestall contaminated IoT units from spreading to different devices. In many cases, the use of computing device gaining knowledge of algorithms to impenetrable IoT units and networks has given promising effects and may want to show to be extraordinarily really helpful for IoT security.

IoT Security Using Machine Learning

In the year 2021, there will be an estimated 24 billion IoT units related to the web [6]. Like most different technologies, the Internet of Things can be exploited by means of hackers and different deviants in order to weaken networks, deny carrier to one's system, steal treasured information, or habits different malicious activities. As the quantity of IoT gadgets have grown, assaults that make the most susceptible IoT gadgets have elevated exponentially [7]. This makes the safety for the Internet of Things extraordinarily essential in order to guard the indispensable structures that they are linked to, in addition to the data that runs thru them. Recently, the use of a range of computer gaining knowledge of strategies in the subject of IoT safety has been explored with dreams like, countering man-in-the-middle or selective forwarding assaults [8]. Many of these strategies show promising with excessive prediction accuracies, self- gaining knowledge of models, real-time security, and expanded efficiency.

The time period "The Internet of Things" was once created in 1999 by using Kevin Ashton [9], who used the phrase to describe the use of RFID technological know-how in the provide chain for Procter & Gamble. Nowadays, the time period is used in a broader sense, describing each object, or thing, that incorporates sensors, software, and different hardware which are used to join to and speak with the internet. So far, severa businesses, homes, cities, and different groups have deployed person networks of IoT gadgets in order to accomplish tasks and dreams with higher effectivity such as, managing strength consumption, calling the authorities in an emergency, or finishing duties at domestic with voice commands. One promising utilization of an IoT community as phase of a city's improvement can be determined in the metropolis of Padova. They make use of IoT gadgets to screen avenue lighting, carbon monoxide levels, noise levels, and greater [10]. Although these networks of IoT gadgets can enhance our general of living, they, like most different technology, can be exploited by means of hackers and criminals if they are now not included correctly.

The safety of IoT units is one of the most vital issues for the Internet of Things that wishes to be addressed. New methods to impenetrable IoT units are in regular improvement in order to higher guard towards cyberattacks on IoT gadgets and networks, which are equally ever-changing and evolving. However, growing safety measures for IoT gadgets has its very own set of challenges. For one, many IoT gadgets function below the constraint of low energy and, therefore, function with a low computing electricity as nicely [11]. Raza et al. have completed lookup about the use of the limited software protocol (CoAP), which may want to be a fee fantastic answer to guard the switch of facts for IoT structures that are limited by way of computing strength making real-time protection safety unachievable [9]. In addition, now not all IoT gadgets are of comparable structure, which makes vulnerabilities and bugs difficult to music [11]. Authors have accomplished lookup on the use of blockchaining, a approach that has been used in conjunction with the cryptocurrency, Bitcoin [10]. They have brought blockchaining to an IoT machine in order to create a protection shape that will embody and shield a household's listing of IoT devices, however it would require every family to function their personal non-public blockchain. Another set of protection measures for cell units and the cloud surroundings is the use of biometric authentication techniques the use of desktop studying algorithms. The use of biometric authentication strategies for cell units related to IoT networks radically reduces the danger of facts being stolen and helps make certain that a individual having access to some private piece of data is the supposed person [11][12][13].Ultimately, IoT cybersecurity is nevertheless missing for today's protection panorama when in contrast to different areas with remarkable protection like anti-fraud structures for on-line payments. IoT gadgets nevertheless face challenges in areas of protection along with Malware Detection and Prevention, and Object Identification. So far, options to these areas have been underdeveloped in the area of IoT structures [8]. In addition, the variety of botnet assaults the usage of compromised IoT gadgets is growing significantly. In September 2016, the internet site of a pc safety consulting company used to be hit with 620 Gbps of visitors from an IoT botnet. At the equal time, an even greater Distributed Denial of Service (DDoS) attack, the usage of Mirai malware, peaked at 1.1 Tbps and centered the website hosting cloud provider company OVH [14]. Attacks like these are solely getting extra common and, as the variety of IoT gadgets increase, the volumetric information of botnet assaults develop significantly, making it more difficult for web- infrastructure and website- security groups like Cloudflare to mitigate assaults [15]. Due to the developing occurrence of IoT attacks, there is a want to locate safety measures that will defend IoT gadgets from being exploited.
Authentication Schemes using Machine Learning

      Behavioral biometrics encompasses a extensive range of points that can be studied and used for authentication purposes. Behavioral Biometrics is an implicit way to test the conduct of the user, which is something that can't be without difficulty copied. Checking the behavioral trait as an alternative of a bodily trait eliminates the issues requiring more hard- ware for scanners and the reliance on environmental factors, making the gadget greater environment friendly and extra secure. Behavioral biometrics can consist of something from a user's behaviors, such as the customers contact patterns, keystroke dynamics, and mouse dynamics [10-12]. Most consumer authentication schemes the usage of behavioral biometrics observe the equal technique for their methodology. This consists of the steps of accumulating the data, extracting the behavioral biometric features, the usage of Machine Learning algorithms for classification, and then the use of these algorithms to create a selection on if the facts can be related to the licensed user. The facts is accumulated both implicitly by means of an utility or scanner on the user's device, the particular behavioral elements the authentication gadget is reading are then extracted from the information collected. From there, Machine Learning algorithms take in the extracted aspects and evaluate them to correct samples of the user's behaviors. In [13], authors created an software referred to as Touchstroke that implicitly video display units the hand micro-movements and touch stroke patterns of the user. However, it was once discovered in this article that these behavioral facets did no longer function properly whilst customers had been in motion, accordingly, decreasing the accuracy of the application. This problem with accumulating behavioral records whilst the consumer was once in action used to be observed to be an problem in different articles as properly [14-16] and [5]. Also, many research do now not reflect on consideration on how a user's behaviors can alternate over time, such as the consumer getting old and typing slower, the consumer now not being in a position to use their dominant hand, and different long-term behavioral modifications [14]. This survey pursuits to consider a couple of research related to behavioral-based biometric structures and talk about which points work best, and which elements want improvements via future research.

      Another feasible technique of dynamic authentication is PLA. PLA is the act of authenticating a person primarily based on the bodily attributes of a acquired sign from stated consumer in a network. The variance of the bodily attribute used for authentication can be observed in. As hostile to biometric-based authentication, PLA makes use of bodily attributes of a acquired sign to authenticate a consumer except the want for an awful lot additional equipment. With wired and wi-fi networks' hastily developing significance due to the expanded use of the Internet of Things (IoT) devices, as properly as the current developments in 5G networks, the safety of these networks are of the utmost significance in making sure their extensive implementation and reliability. A frequent instance of the use of bodily layer authentication in a Multiple Input Multiple Output (MIMO) net- work is the Bob, Alice, and Eve example. This popular scheme is most used throughout the literature. Bob and Alice are relied on customers on a MIMO community sending data between themselves. Bob is most oftentimes receiving a sign from Alice. Eve is a nefarious user, making an attempt to spoof as one of the relied on customers through manipulating their very own

signal's bodily attributes. It is then up to Bob to authenticate every sign he receives and decide if it is virtually from Alice or a spoofed sign from Eve. PLA techniques have been proposed to high-quality mitigate the opportunity of unintentional authentication. Addition- ally, PLA can additionally be used to stop malicious assaults in opposition to a network, such as a Denial of Service (DOS) assault or man-in-the-middle attacks. As the common social, economic, and private dependence on wi-fi networks continues to increase, the funding in the reliability and protection of these networks is greater than worthwhile.

## II. CONCLUSION

Through the assessment of these surveys, it used to be observed that Touchstroke conduct biometrics and the computing device studying algorithm of random woodland carried out higher than their behavioral biometric and computer mastering counterparts. Touchstroke biometrics, as viewed by way of the lookup accumulated from different articles, is without problems the most invulnerable and fine behavioral biometric to be used for consumer authentication. Also, the use of behavioral biometrics in widespread makes the authentication scheme much less expensive for each the producers to produce and the customers to purchase. Also mentioned in this survey was once the use of different Machine Learning algorithms in behavioral biometric primarily based consumer authentication. While all algorithms carried out properly in their respective studies, Random Forest was once regularly the first-class performing algorithm.

This study is about gaining knowledge of algorithms in
IoT safety via reviewing various articles
which carried out a variety of strategies and strategies in order
to locate options for some of the issues in IoT security. From the beginning, this learn about highlighted the enormity of the Internet of Things, as nicely as its viable to be exploited. Then, it started to center of attention on the man or woman instances in which the use of computing device studying may
also advantage IoT protection together with its use for malware and intrusion detection and the identification of unknown IoT devices. Many of these techniques that had
been featured both concluded that Random Forest was once the exceptional computer gaining knowledge of algorithm for their strategies or had been the usage of Random Forest on its very own from the get-go. However, others located that via the use of a couple of computer gaining knowledge of algorithms the accuracy and precision would improve. One algorithm's weaknesses would be complimented via the different algorithm, which produced greater accuracies than a single algorithm ought to acquire on its own.